\documentclass[aps,pre,twocolumn,superscriptaddress]{revtex4}
\usepackage[dvips]{graphics}
\usepackage{graphicx}
\usepackage{amsfonts}
\usepackage{amssymb}
\usepackage{amsmath}
\usepackage{subfigure}

\begin{document}

\title{Mobility of Discrete Solitons in Quadratically Nonlinear Media}
\author{H. Susanto}
\affiliation{Department of Mathematics and Statistics, University of Massachusetts,
Amherst MA 01003-4515, USA}
\author{P. G.\ Kevrekidis}
\affiliation{Department of Mathematics and Statistics, University of Massachusetts,
Amherst MA 01003-4515, USA}
\author{R.\ Carretero-Gonz\'alez}
\affiliation{Nonlinear Dynamical Systems Group, Department of Mathematics and Statistics,
and Computational Science Research Center, San Diego State University, San
Diego CA, 92182-7720, USA}
\author{Boris A.\ Malomed}
\affiliation{Department of Interdisciplinary Studies, School of Electrical Engineering,
Faculty of Engineering, Tel Aviv University, Tel Aviv 69978, Israel}
\author{D. J.\ Frantzeskakis}
\affiliation{Department of Physics, University of Athens, Panepistimiopolis, Zografos,
Athens 15784, Greece }

\begin{abstract}
We study the mobility of solitons in second-harmonic-generating lattices.
Contrary to what is known for their cubic counterparts, discrete quadratic
solitons are mobile not only in the one-dimensional (1D) setting, but also
in two dimensions (2D). We identify parametric regions where an initial kick
applied to a soliton leads to three possible outcomes, namely, staying put,
persistent motion, or destruction. For the 2D lattice, it is found that, for
the solitary waves, the direction along which they can sustain
the largest kick and can attain
the largest speed is the diagonal. 
Basic dynamical properties of the discrete solitons are
also discussed in the context of an analytical approximation, in terms of an
effective Peierls-Nabarro potential in the lattice setting.
\end{abstract}

\maketitle





\textit{Introduction}. In the past several years, tremendous progress has
been observed in studies of nonlinear dynamical systems on lattices \cite%
{reviews}. In a considerable part, this development is fueled by the
continuous expansion of relevant physical applications, including spatial
dynamics of optical beams in waveguide arrays \cite{reviews1}, temporal
evolution of Bose-Einstein condensates in deep optical lattices \cite%
{reviews2}, transformations of the DNA double strand \cite{reviews3}, and so
on.

A ubiquitous dynamical-lattice system is\ represented by the discrete
nonlinear Schr{\"{o}}dinger equation \cite{christo,reviews,reviews1} with
cubic ($\chi ^{(3)}$) nonlinearity. It has been used to model a variety of
experiments featuring, among others, formation of discrete solitons, lattice
modulational instability, buildup of the Peierls-Nabarro (PN)\ barrier
impeding the soliton motion, diffraction management, and soliton
interactions \cite{eise1,eise2,eise3,florence,steg1,steg2}.

Substantial activity has also been aimed at lattices with quadratic ($\chi
^{(2)}$) nonlinearity, boosted, in particular, by the recent experimental
realization of discrete $\chi ^{(2)}$ solitons in nonlinear optics \cite%
{steg3}. A variety of topics have been studied for such media both
theoretically and experimentally, including the formation of 1D and 2D
solitons \cite{mal0,dkl,us} (see also reviews \cite{buryak,steg4}),
observation of the modulational instability in periodically poled lithium
niobate waveguide arrays, finite few-site lattices \cite{old}, $\chi ^{(2)}$
photonic crystals \cite{sukho}, cavity solitons \cite{pertsch}, and
multi-color localized modes \cite{vic0}. In addition, the same lattice
models with the quadratic nonlinearity may be used to describe the dynamics
of Fermi-resonance interface modes in multilayered systems based on organic
crystals \cite{agran}. A variety of solutions have been obtained in the
latter context \cite{agran1}.

A fundamental difference of $\chi ^{(2)}$ continua from their $\chi ^{(3)}$
counterparts \cite{sulem} is that they feature no collapse in 2D and 3D
cases \cite{KR}, which paves the way to 
create stable 2D \cite{steblina} and
3D \cite{HaoHe} quadratic solitons. On the other hand, due to the presence
of collapse-type phenomena  in 2D and 3D $\chi ^{(3)}$ continua, lattice
solitons in the corresponding discrete setting are subject to \textit{%
quasi-collapse}. As a consequence, they only exist with a norm (``mass'')
exceeding a certain threshold value \cite{wein}, and they are strongly
localized (on few lattice sites), hence 2D and 3D $\chi ^{(3)}$ solitons are
strongly pinned to the lattice and cannot be motile \cite{vicencio}.

The absence of the trend to the catastrophic self-compression in the 2D $%
\chi ^{(2)}$ medium suggests that the corresponding lattice solitons may be
broad and therefore \emph{mobile}, being loosely bound to the lattice. The
aim of this work is to investigate the mobility of 1D and, especially, 2D
solitons in $\chi ^{(2)}$ lattices. Besides its importance for the use of
such waves in photonic applications, the topic is of fundamental interest by
itself, as it will reveal a family of \emph{mobile solitons} in 2D lattices.
Thus far, the only example of mobility was provided by solitons in a 2D
lattice with saturable nonlinearity \cite{vic0} (the \textit{%
Vinetskii-Kukhtarev} model \cite{ViKu}, in which the mobility of 1D solitons
was examined 
in Ref. \cite{Yugo}). In this work, we identify parametric regions of stable
motion of $\chi ^{(2)}$ solitons on 1D and 2D lattices and, for the first
time in the 2D case, we study \emph{anisotropy} of the mobility of 2D
lattice solitons (for propagation off of the principal directions of the
lattice). First, we will introduce the model and develop an analytical
approach to solitary waves and the respective PN barrier in $\chi ^{(2)}$
lattices. Then, systematic numerical results for the soliton mobility in 1D
and 2D lattices will be reported.

\textit{The model and analytical results}. Following Ref. \cite{us}, we
introduce a system including equations for the fundamental-frequency (FF)
and second-harmonic (SH) waves, $\psi _{m,n}(t)$ and $\phi _{m,n}(t)$. In
the 2D setting the model has the following form (its 1D counterpart will
also be used below):
\begin{eqnarray}
i\frac{d}{dt}\psi _{m,n} &=&-\left( C_{1}\Delta _{2}\psi _{m,n}+\psi
_{m,n}^{\star }\phi _{m,n}\right) \,,  \label{geq3} \\
i\frac{d}{dt}\phi _{m,n} &=&-\frac{1}{2}\left( C_{2}\Delta _{2}\phi
_{m,n}+\psi _{m,n}^{2}+k\phi _{m,n}\right) \,,  \label{geq4}
\end{eqnarray}%
where $\Delta _{2}u_{m,n}\equiv
u_{m+1,n}+u_{m-1,n}+u_{m,n+1}+u_{m,n-1}-4u_{mn}$ is the discrete Laplacian, $%
C_{1}$ and $C_{2}$ are the FF and SH lattice-coupling constants, and $k$ is
the mismatch parameter. Equations (\ref{geq3}) and (\ref{geq4}) conserve the
corresponding Hamiltonian and the Manley-Rowe (MR) invariant, $%
I=\sum_{m,n}\left( \left\vert \psi _{m,n}\right\vert ^{2}+2\left\vert \phi
_{m,n}\right\vert ^{2}\right) $.

Stationary solutions are looked for as $\{\psi _{m,n}(t),\phi
_{mn}(t)\}=\{e^{-i\omega t}\Psi _{m,n},e^{-2i\omega t}\Phi _{m,n}\}$, where
distributions $\Psi _{m,n},\Phi _{m,n}$ are real for fundamental solitons,
and may be complex for more elaborate patterns, such as vortices \cite{us}.
To set discrete solitons in motion, one must overcome the above-mentioned PN
barrier, i.e., the energy difference between static solitons, centered,
respectively, on a lattice site and between sites. To obtain an approximate
analytical expression for the barrier, we consider the continuum limit, in
which stationary functions $\Psi $ and $\Phi $ depend only on the radial
variable which is the continuum limit of $r\equiv \sqrt{\left(
m^{2}+n^{2}\right) /C_{1}}$ and obey equations%
\begin{eqnarray}
\omega \Psi +\left( \Psi ^{\prime \prime }+\Psi ^{\prime }/r\right) +\Psi
\Phi  &=&0\,,  \notag \\
\left( 4\omega +k\right) \Phi +C\left( \Phi ^{\prime \prime }+\Phi ^{\prime
}/r\right) +\Psi ^{2} &=&0,  \label{stat}
\end{eqnarray}%
where $C\equiv C_{2}/C_{1}$, and the prime stands for $d/dr$. In this limit,
the soliton is approximated by the following ansatz with amplitudes $A$ and $%
B$:
\begin{eqnarray}
\left\{
\begin{array}{c}
\Psi  \\
\Phi
\end{array}%
\right\}  &=&\left\{
\begin{array}{c}
A \\
B/\sqrt{2}%
\end{array}%
\right\} \sqrt{\frac{\sinh \left( 2\sqrt{\left\{ |\omega |,|\chi |\right\} }%
r\right) }{\sqrt{\left\{ |\omega |,|\chi |\right\} }r}}  \notag \\
&&\mathrm{sech}\left( 2\sqrt{\left\{ |\omega |,|\chi |\right\} }r\right) ,
\label{ansatz2D}
\end{eqnarray}%
where $\omega $ and $\chi \equiv \left( 4\omega +k\right) /C$ must be
negative. These expressions have the correct 2D asymptotic forms at $r$\-$%
\rightarrow \infty $, $\left\{ \Psi ,\Phi \right\} \sim r^{-1/2}\exp \left( -%
\sqrt{|\left\{ \omega ,\chi \right\} }r\right) $. Substituting the ansatz in
Eqs. (\ref{stat}) and demanding its validity at $r\rightarrow 0$, we obtain $%
B=(23/3)|\omega |,~A=(23/3)\sqrt{\omega \left( 4\omega +k\right) }$.

The Hamiltonian corresponding to the axially symmetric real solutions of the
continuum equations is
\begin{eqnarray}
H &=&\pi \int_{0}^{\infty }rdr\left[ 2\left( \Psi _{r}^{\prime }\right)
^{2}+C\left( \Phi _{r}^{\prime }\right) ^{2}-2\Phi \Psi ^{2}\right.   \notag
\\
\left. -k\Phi ^{2}\right]  &=&\pi \int_{0}^{\infty }rdr\left( 2\omega \Psi
^{2}+4\omega \Phi ^{2}+\Phi \Psi ^{2}\right) ,  \label{simple}
\end{eqnarray}%
where the derivatives were eliminated using integration by parts and Eqs. (%
\ref{stat}). To derive the PN potential, we apply the lattice discretization
to final expression (\ref{simple}) by defining $H_{\mathrm{latt}}=\frac{1}{2}%
\int \int \left( 2\omega \Psi ^{2}+4\omega \Phi ^{2}+\Phi \Psi ^{2}\right)
\mathrm{gr}(x,y)~dxdy$, where the grid function is%
\begin{equation}
\mathrm{gr}(x,y)\equiv \!\!\sum_{m,n=-\infty }^{+\infty }\!\!\delta \left(
x-m\right) \delta \left( y-n\right) =\!\!\sum_{p,q=-\infty }^{+\infty
}\!\!e^{2\pi i\left( px+qy\right) }.  \label{Grid}
\end{equation}%
In the quasi-continuum approximation (which implies small $|\omega |$ and $%
|\chi |$), the leading terms in $H_{\mathrm{latt}}$ correspond to $\left(
p,q\right) =\left( \pm 1,0\right) $ and $\left( 0,\pm 1\right) $ and yield
the PN potential with an exponentially small amplitude, $U=U_{0}\left[ \cos
\left( 2\pi \xi \right) +\cos \left( 2\pi \eta \right) \right] $, where $%
(\xi ,\eta )$ are the coordinates of the soliton's center. An expression for
the
amplitude $U_{0}$ is simplest in the case of $|\chi |>2|\omega |$, which
corresponds to numerical results presented below (with $\omega =-0.25$, $%
\chi =-0.75$); in this case, the second term in $H_{\mathrm{latt}}$
dominates. Fitting the slowly varying part of the integrand to a Gaussian,
we thus obtain $U_{0}=-\alpha \left( |\omega |^{3}/|\chi |\right) \exp
\left( -3\pi ^{2}/(10|\chi |)\right) $, with $\alpha \equiv (2\pi
/15)23^{2}\approx 222$.

\textit{Numerical Results.} In the 1D and 2D cases alike, we used lattices
with periodic boundary conditions, in order to allow indefinitely long
progressive motion of solitons. First, we found standing lattice-soliton
solutions $\left\{ \Psi _{m,n}^{(0)},\Phi _{m,n}^{(0)}\right\} $, by means
of fixed-point iterations. Next, dynamical simulations were initialized by
applying a \textit{shove} (kick) to those solutions, which corresponds to
initial conditions%
\begin{equation}
\left\{
\begin{array}{c}
\psi _{m,n} \\
\phi _{m,n}%
\end{array}%
\right\} =e^{i\left( S/C_{1,2}\right) \left( m\cos \theta +n\sin \theta
\right) }\left\{
\begin{array}{c}
\Psi _{m,n}^{(0)} \\
\Phi _{m,n}^{(0)}%
\end{array}%
\right\} ,  \label{shove}
\end{equation}%
where $S$ and $\theta $ determine the size and orientation of the shove
vector.

\begin{figure}[tb]
\centerline{
\includegraphics[height=5cm,width=4.5cm,angle=0,clip]{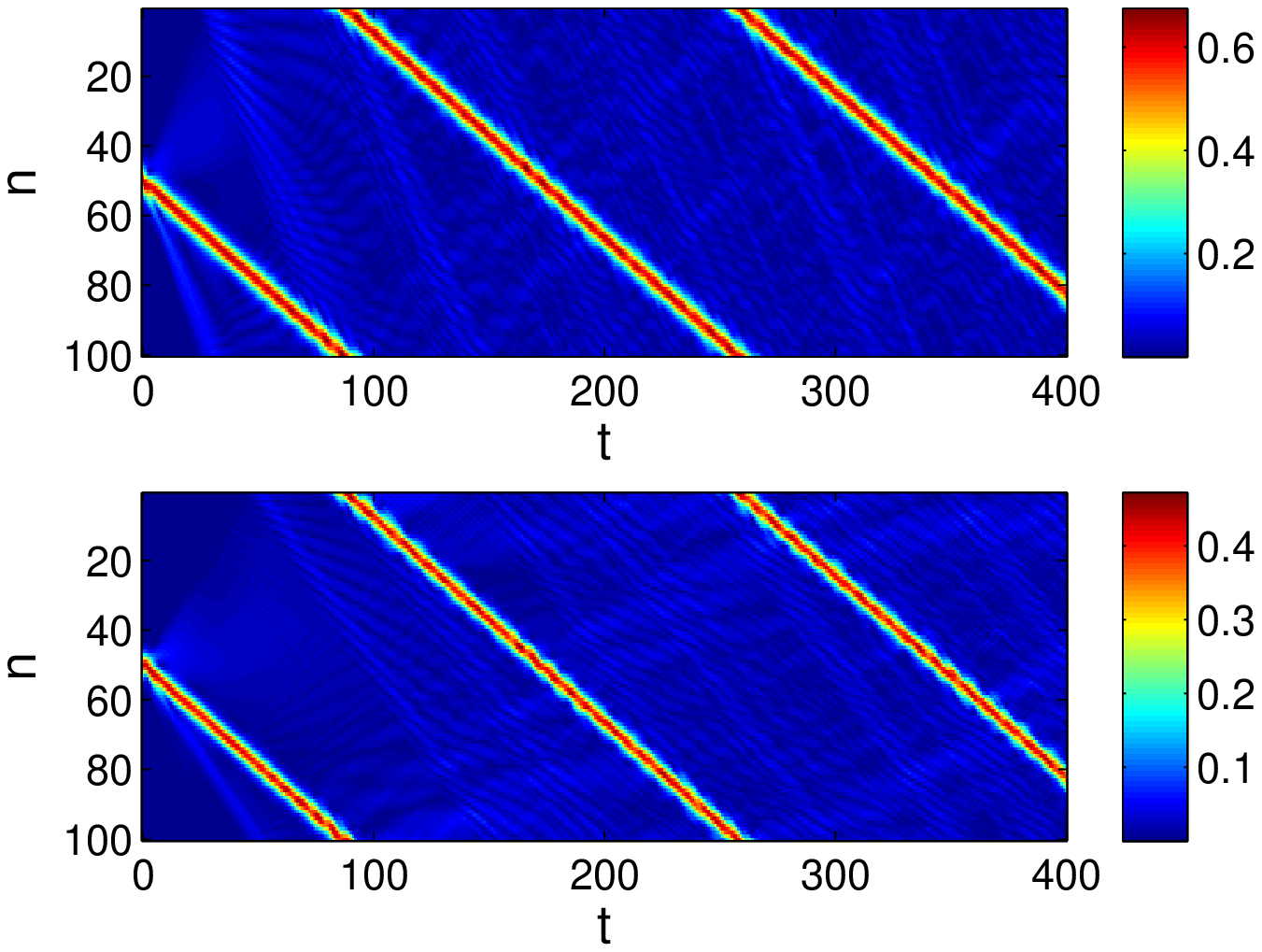}
\includegraphics[height=5cm,width=4.5cm,angle=0,clip]{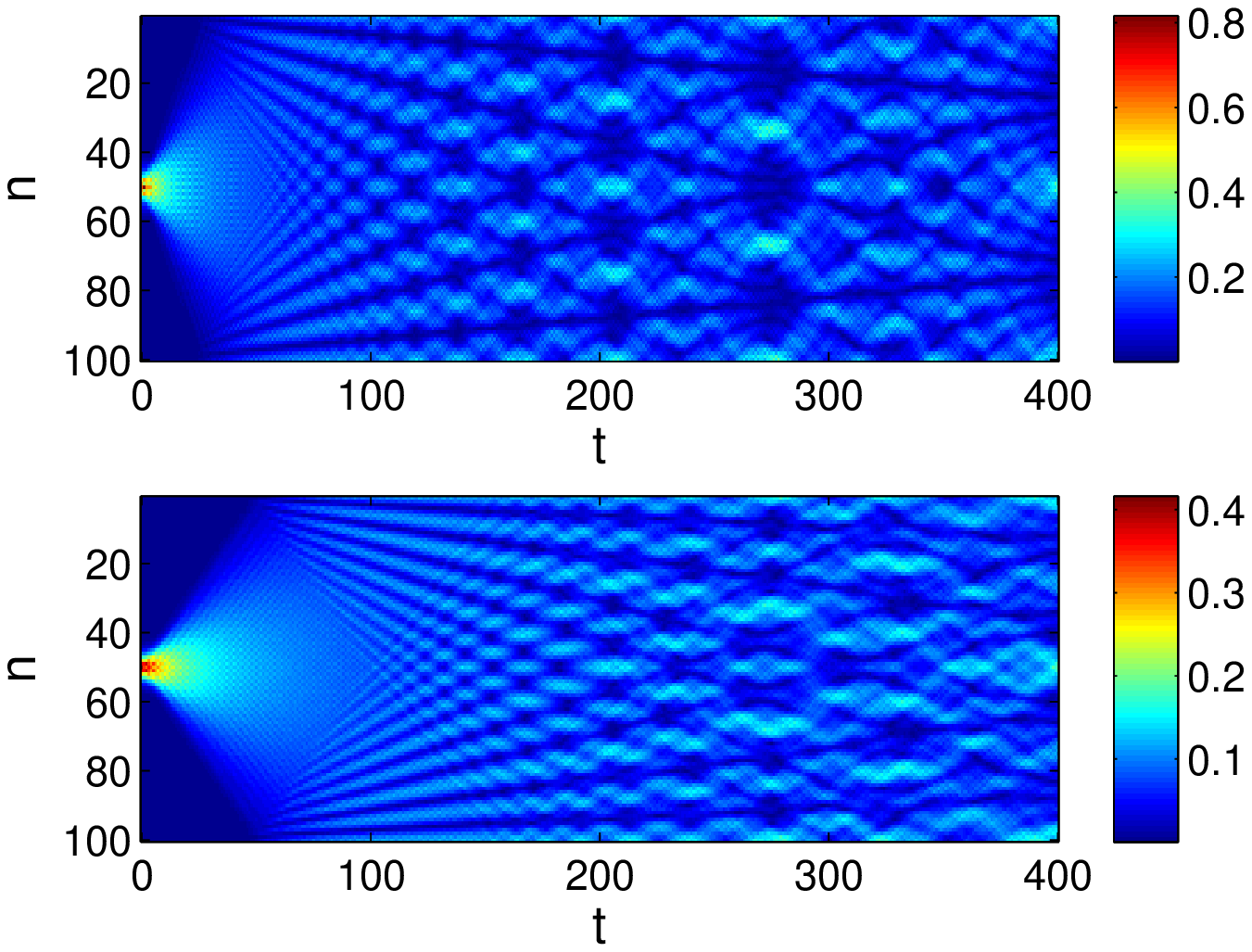}
}
\caption{(Color online) Space-time contour plots of $\left\vert \protect\psi %
_{m,n}\right\vert ^{2}$ and $|\protect\phi _{m,n}|^{2}$ for the FF and SH
fields (top and bottom panels) in the 1D lattice with periodic boundary
conditions, for $C_{1}=C_{2}=1,\,\protect\omega =-0.25,\,k=0.25,$ and the
shove's strength $S=0.4$ and $3.0$ (left and right panels
respectively). The boosted
soliton sets in stable motion in the former case, and is destroyed in the
latter case.}
\label{Fig1}
\end{figure}

\begin{figure}[tb]
\centerline{
\includegraphics[width=7.cm,angle=0,clip]{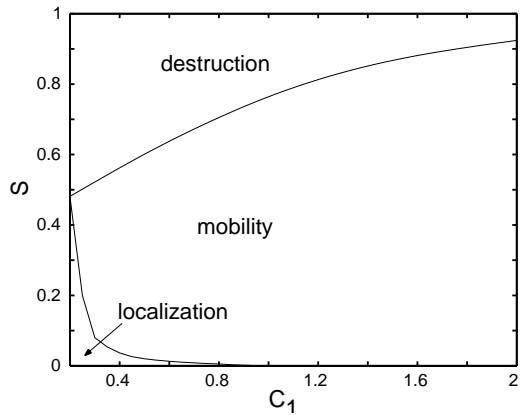}
}
\caption{A diagram in the plane of the coupling strength, 
$C_{1}=C_{2}$, and shove's strength, $S$, showing different outcomes of
kicking the quiescent soliton in the 1D lattice, for $\protect\omega =-0.25$
and $k=0.25$. (``localization" means that the soliton remains quiescent). }
\label{Fig2}
\end{figure}

Examples of stable motion and destruction of the 1D lattice
soliton subjected to the shove are displayed in Fig.~\ref{Fig1},
and systematic results, obtained with variation of $S$ and
$C_{1}=C_{2}$, are summarized in Fig.~\ref{Fig2}. The
destruction was registered as the outcome if the kicked soliton
would eventually lose more than $30\%$ of the initial value of its
MR invariant. For the coupling strength such as that corresponding
to Fig.~\ref{Fig1}, there are, practically, only two outcomes,
\textit{mobile waves} and \textit{wave destruction,} observed with
different initial kicks. However, for weaker couplings (i.e.,
stronger discreteness), ``localization" is also possible: if the
shove's strength, $S $, is below a lower critical value,
$S_{\mathrm{cr}}^{(0)}$, the soliton survives without acquiring
any velocity. The latter outcome is explained by noting that the
kinetic energy, $E_{\mathrm{kin}}\sim S^{2}$, imparted to
the soliton by the shove, may be insufficient to overcome the PN barrier ($%
2U_0$ as defined above). Because $U_{0}$ decays exponentially
with the increase of the intersite coupling, the
``localization" region in Fig.~\ref{Fig2} is very small. General
features of the 1D situation are: (i) for
$S<S_{\mathrm{cr}}^{(0)}$, the soliton remains quiescent; (ii) for
$S_{\mathrm{cr}}^{(0)}<S<S_{\mathrm{cr}}$, the soliton sets in the
state of persistent propagation; (iii) for $S>S_{\mathrm{cr}}$, the
soliton is destroyed.

\begin{figure}[tb]
\centerline{
~~
\includegraphics[width=4.75cm,height=3.0cm,angle=0,clip]{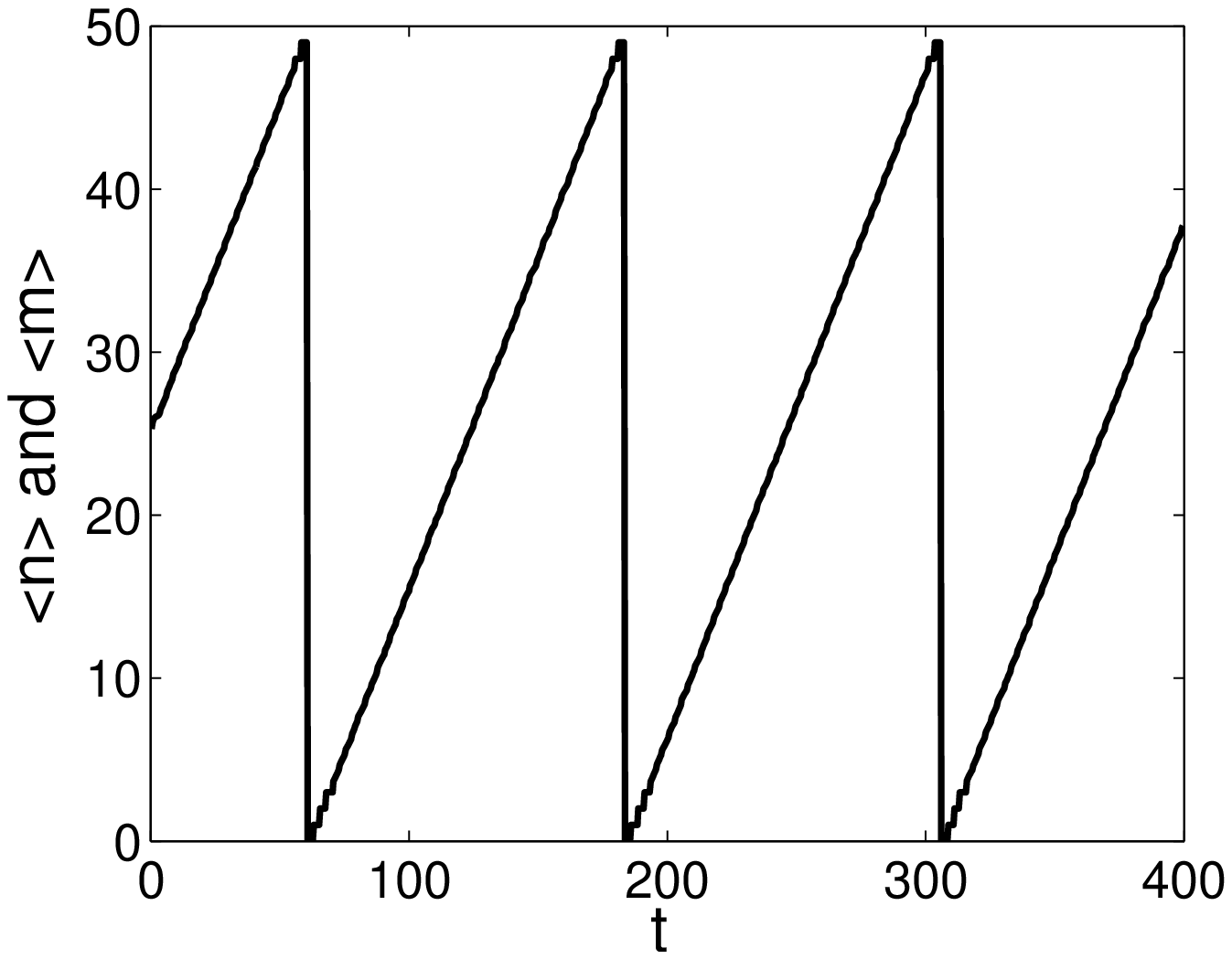}
\hskip-0.5cm
\includegraphics[width=4.75cm,height=3.0cm,angle=0,clip]{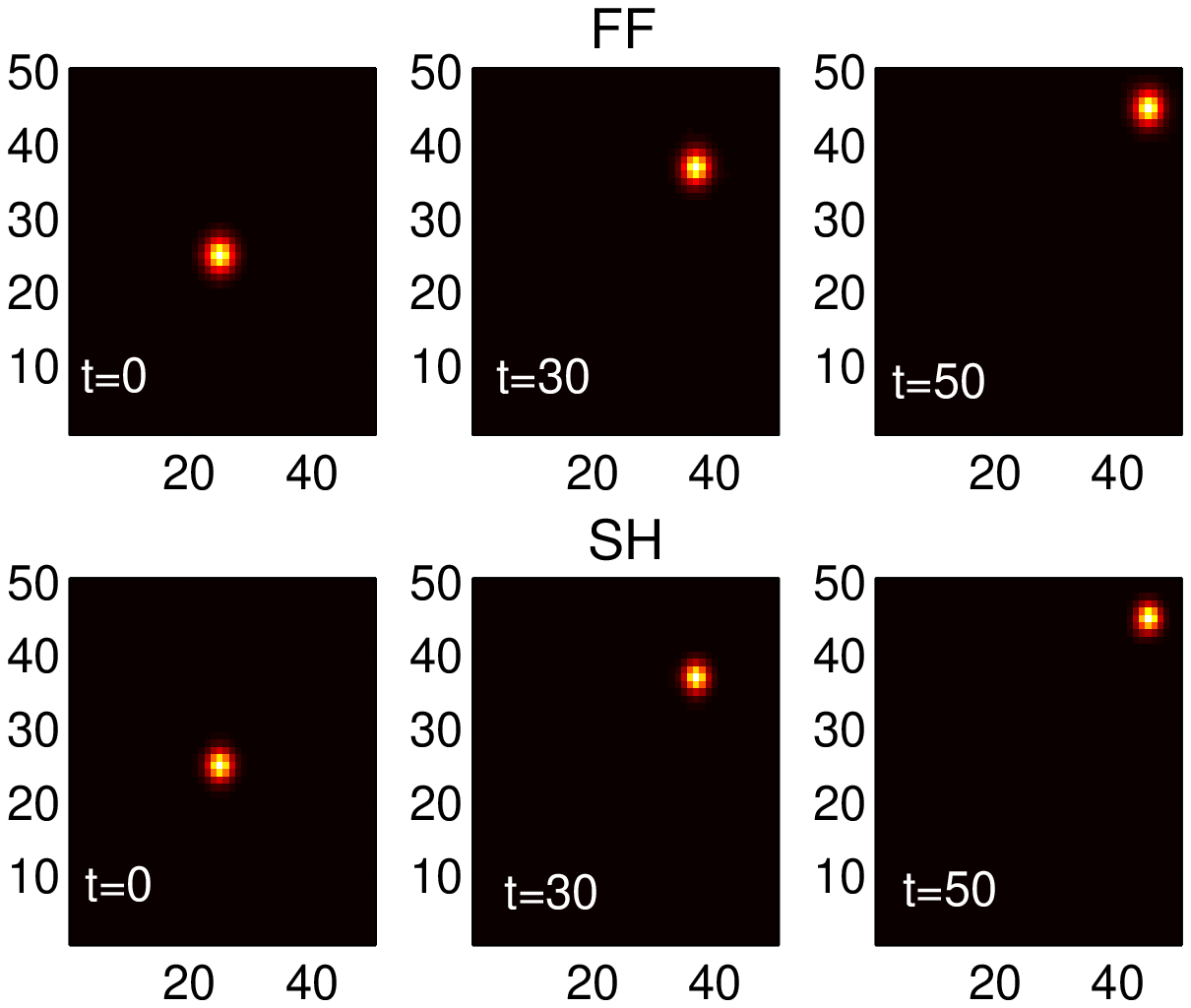}
} 
\vskip0.0cm
\centerline{
~~
\includegraphics[width=4.75cm,height=3.0cm,angle=0,clip]{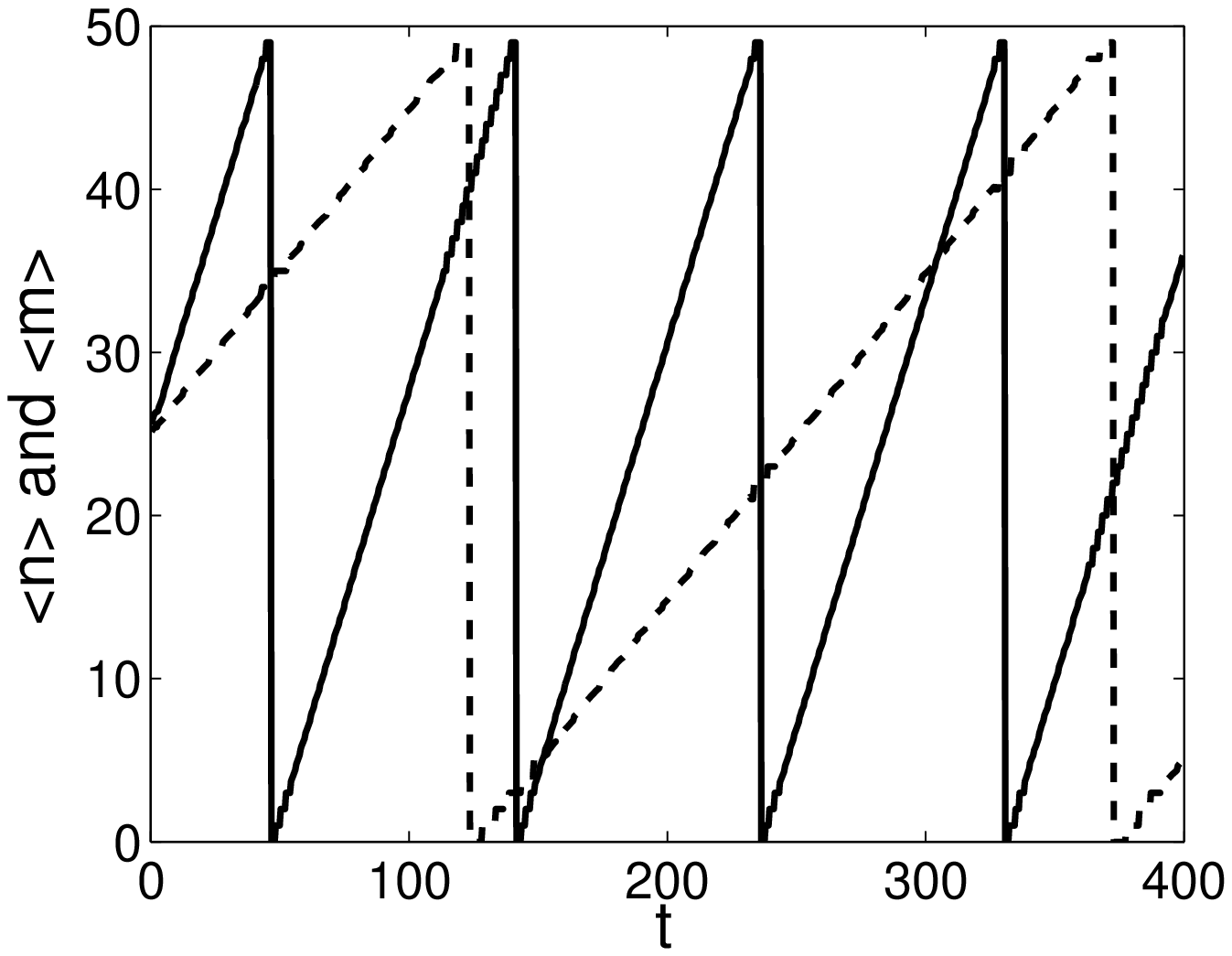}
\hskip-0.5cm
\includegraphics[width=4.75cm,height=3.0cm,angle=0,clip]{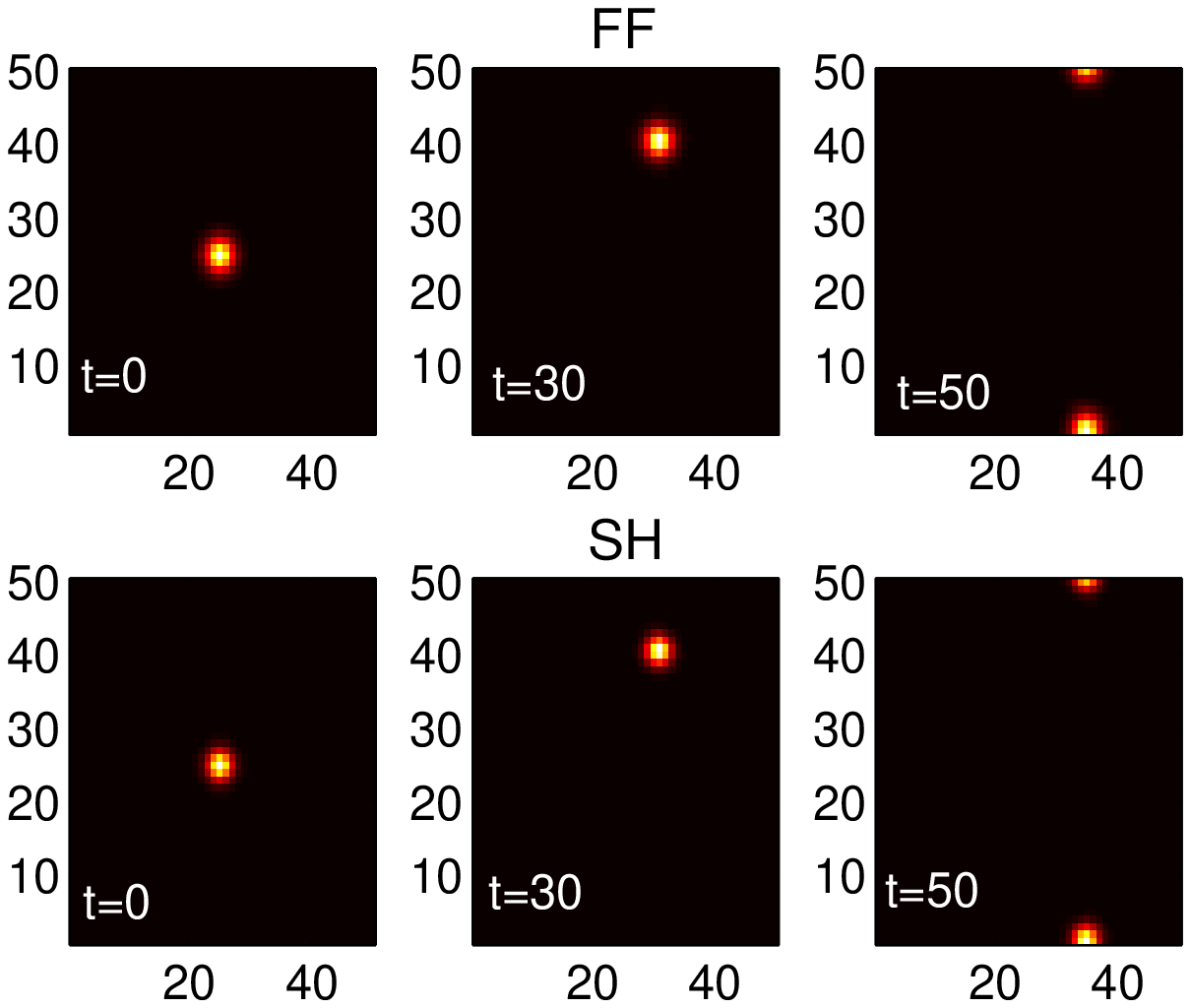}
}
\caption{(Color online) Same as Fig.~\protect\ref{Fig1} and with the same
parameters, but in the 2D periodic lattice, for the propagation in the
diagonal (45 degrees, top panels) and off-diagonal (20 degrees relative to
the lattice bonds, bottom panels) directions. The left panels show
trajectories of the soliton's center, while the right ones display snapshots
of the moving solitons in the FF (top) and SH (bottom) fields at $%
t=0,30,50$.}
\label{Fig3}
\end{figure}

We now turn to the 2D setting, which is more interesting for two reasons.
First, as noted above, in the 2D case the mobility of lattice solitons is a
highly nontrivial feature, practically impossible in the case of the $\chi
^{(3)}$ nonlinearity; second, it is interesting to study anisotropy of the
mobility, i.e., its dependence on the orientation of the initial kick
relative to the (principal directions of the) lattice. Figure \ref{Fig3}
shows two examples of stable moving regimes: one along the lattice
diagonal, and, to our knowledge, the first ever example of the\emph{\ }%
motion on the lattice in an arbitrary direction (neither diagonal, nor along
the bonds).

\begin{figure}[tb]
\centerline{
\includegraphics[width=4.cm,angle=0,clip]{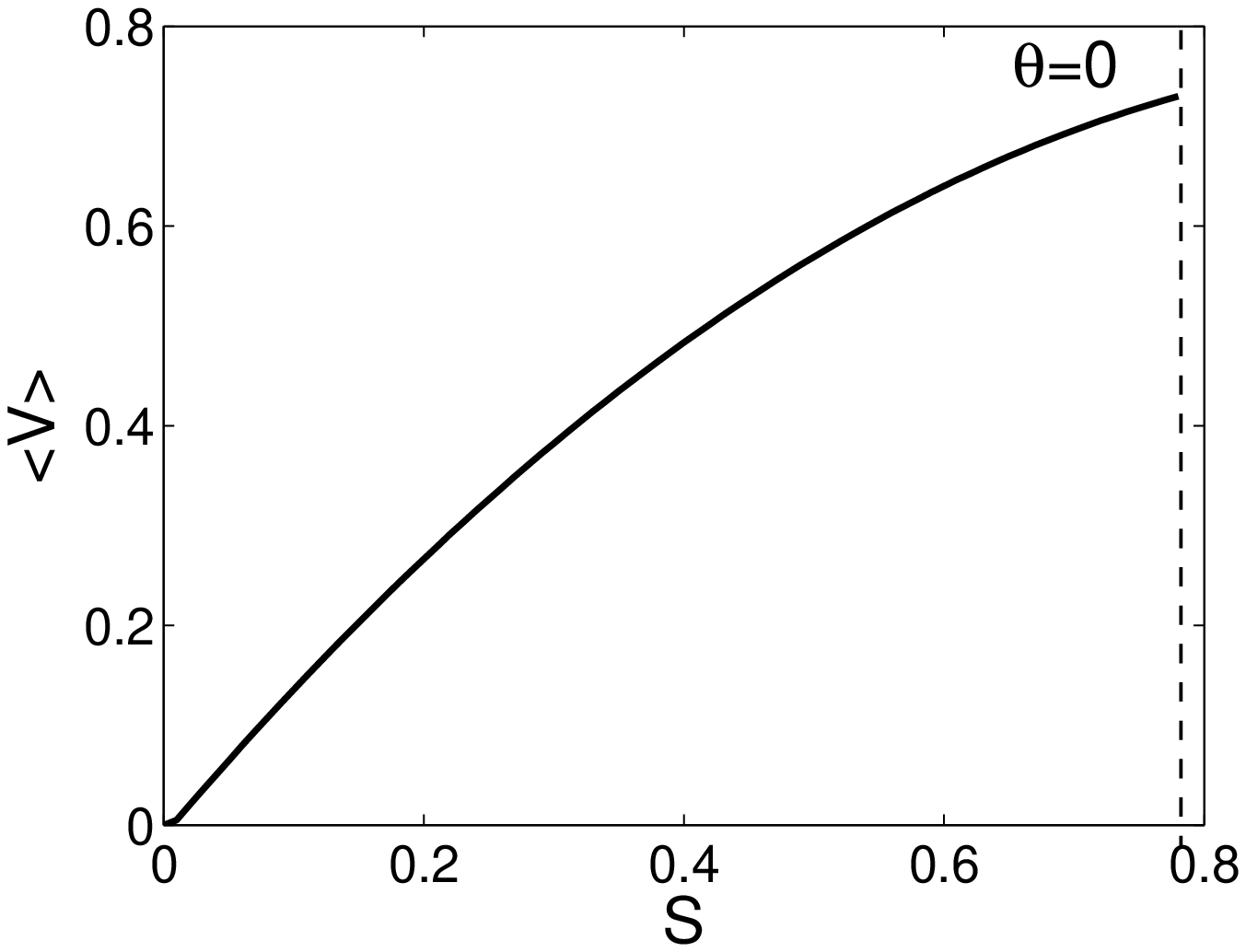}
\includegraphics[width=4.cm,angle=0,clip]{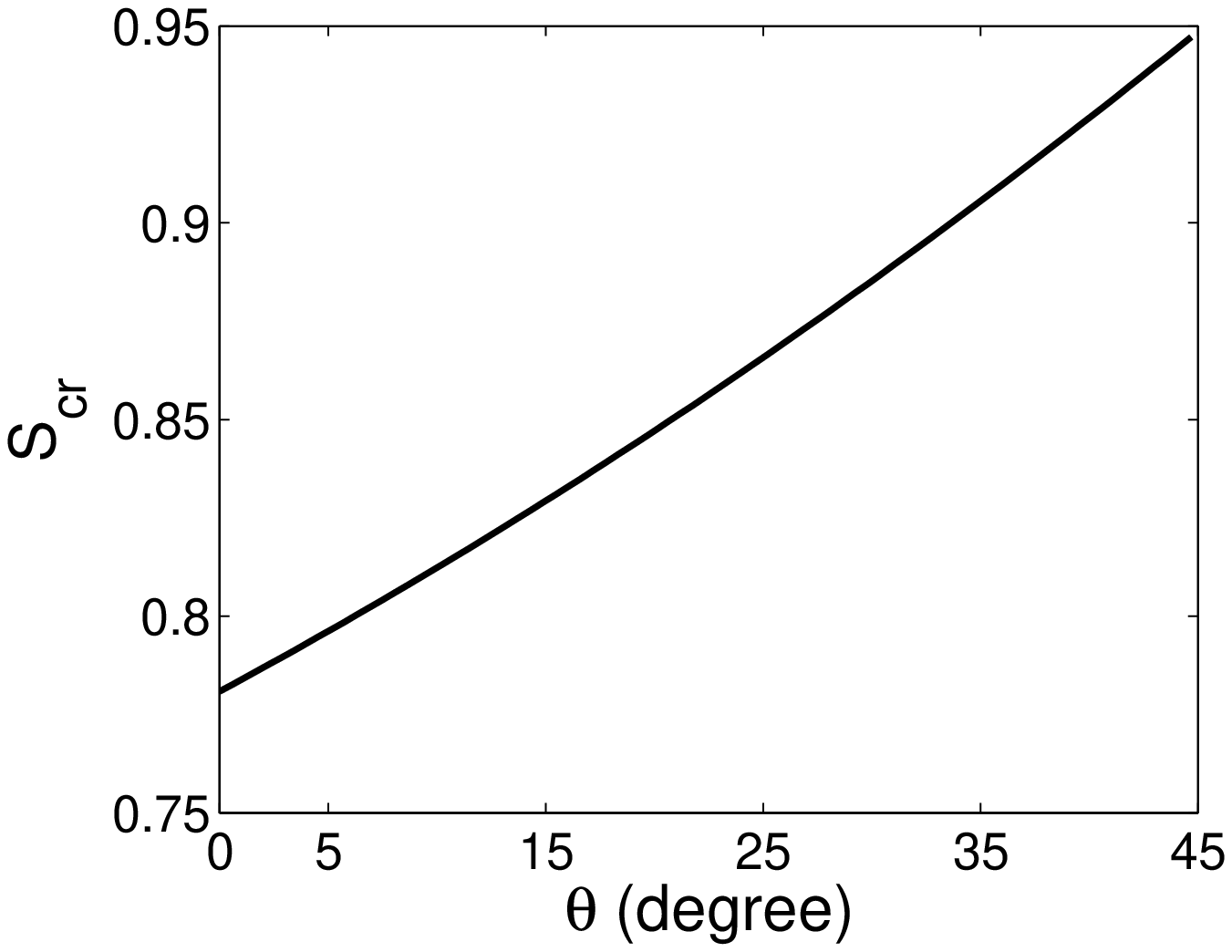}
}
\centerline{
\includegraphics[width=4.cm,angle=0,clip]{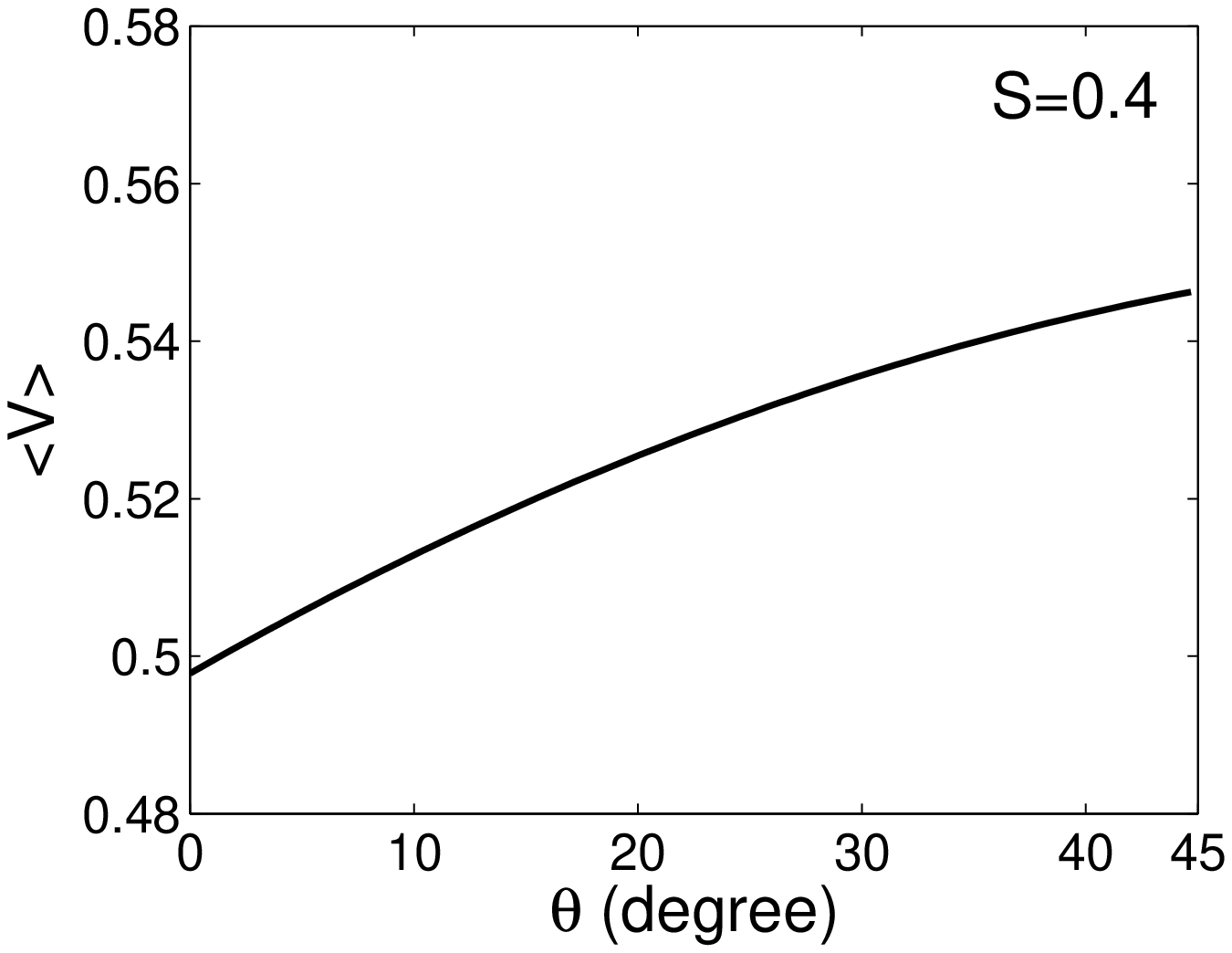}
\includegraphics[width=4.cm,angle=0,clip]{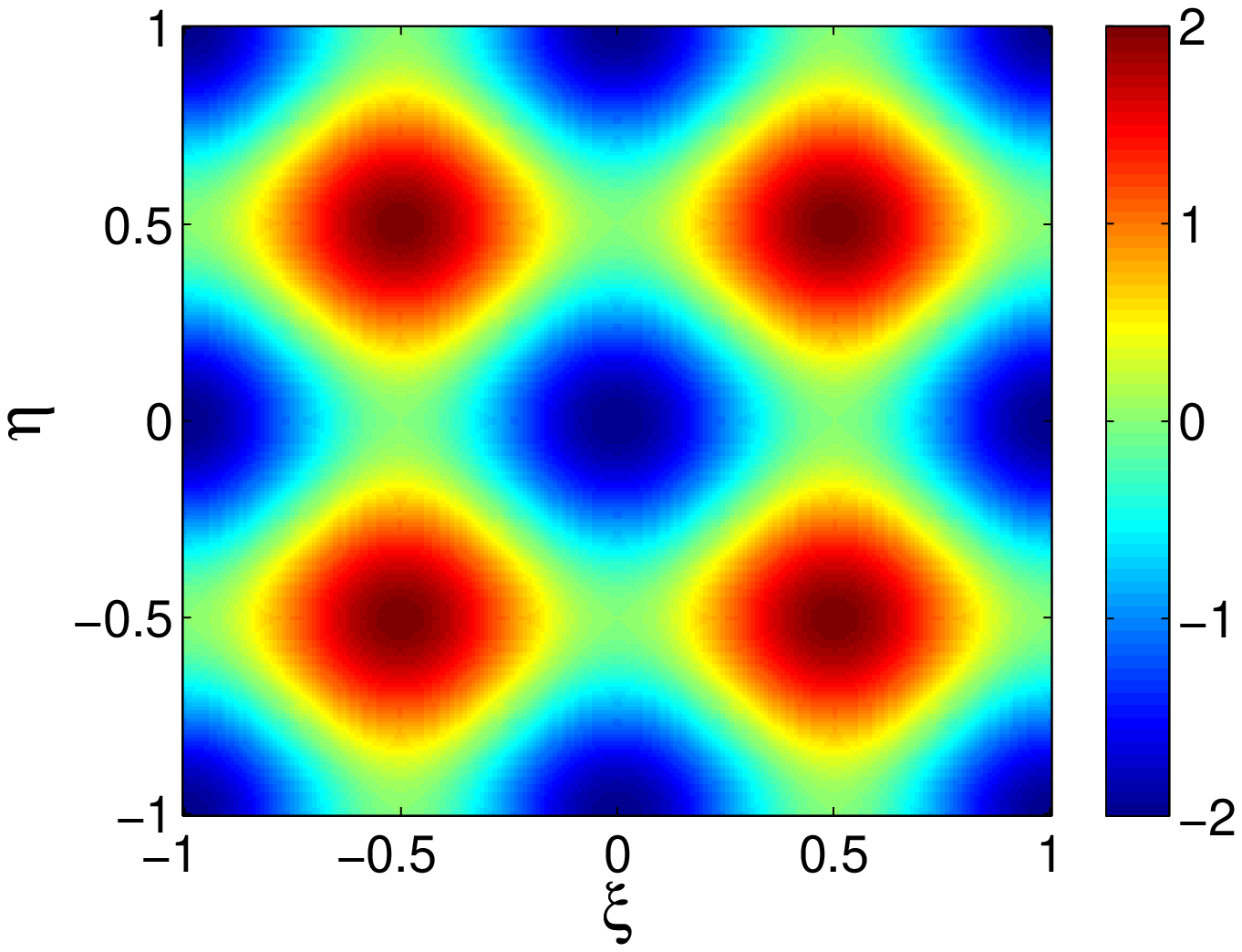}
}
\caption{(Color online) Features of the soliton motion in the 2D periodic
lattice, for $C_{1}=C_{2}=1,\,k=0.25,\,\protect\omega =-0.25$. The
top left panel shows the velocity versus the shove's strength $S$ in the
propagation along the lattice bonds (at angle $\protect\theta =0$); the
vertical dashed line indicates the value of $S_{\mathrm{cr}}$, beyond which
the soliton retains less than $70\%$ of its initial MR invariant, and is
therefore categorized as destroyed. The top right panel depicts $S_{\mathrm{%
cr}}$ as a function of the orientation of the initial kick, $\protect\theta $%
. For a given $S$ ($S=0.4$), the ensuing velocity of the motion is shown
versus $\protect\theta $ in the bottom left corner. In addition, the bottom
right panel shows the analytically predicted PN potential.}
\label{Fig4}
\end{figure}

In Fig.~\ref{Fig4}, we summarize the dependence of the mobility
properties on the strength, $S$, and direction, $\theta $, of the
initial kick. The kicking of the 2D soliton may result in
``localization" (no motion at all), in some interval
$S<S_{\mathrm{cr}}^{(0)}$ ($S<S_{\mathrm{cr}}^{(0)}\approx 0.02 $
in the top left panel of Fig.~\ref{Fig4}). Other generic outcomes
again amount to propagation at a finite velocity, which depends on
$S$, and destruction for a large supercritical kick
$S>S_{\mathrm{cr}}$.

Particularly noteworthy features, specific to the 2D setting, are presented
in the top right and bottom left panels of Fig.~\ref{Fig4}, viz.,
dependences of $S_{\mathrm{cr}}$ and velocity in the moving regime
on $\theta $. These dependences demonstrate that the propagation
direction along which it is easiest to sustain motion (i.e., with 
larger speeds/range of initial kicks) on the square lattice is 
along the diagonal, as the motion in this
direction can be sustained up to larger values of $S_{\mathrm{cr}}$, and is
fastest for given $S$.  Both facts may be qualitatively 
explained by the analytically 
predicted PN potential in the following way. Given the nearest-neighbor
nature of the interactions, in order for the center of the wave to move 
along the diagonal, it has to split along the two lattice directions
and then recombine at the site located diagonally across from the initial
position. The recurrent small-scale symmetric breakups and recombinations
(observed in the numerical data) provide for an effective propagation along
the diagonal direction with a minimum PN barrier (see bottom right panel
of Fig. \ref{Fig4}) and thus result in higher propagation speeds.


We have also examined the situation with $C_{2} < C_{1}$, and
obtained similar results, but with larger $S_{\mathrm{cr}}^{(0)}$. In the
special case of $C_{2}=0$ (no lattice coupling in the SH field), we were not
able to generate moving solitons, which can be easily explained: with $%
C_{2}=0$, Eq. (\ref{geq4}) yields $\Phi _{m,n}=-\left( 4\omega +k\right)
^{-1}\left( \Psi _{m,n}\right) ^{2}$, and the substitution of this in Eq. (%
\ref{geq3}) makes the model equivalent to that with the cubic nonlinearity,
where moving 2D discrete solitons do not exist.

\textit{Conclusions.} In this work, we have examined the mobility of
solitons in 1D and 2D lattices with the quadratic nonlinearity. We have
shown that the solitons feature stable motion much more easily than their
counterparts in 1D lattices with the cubic nonlinearity, and they may also
be mobile on the 2D lattice, where the cubic solitons cannot move at all. In
the 2D lattice, we have for the first time reported a possibility of motion
of the soliton in an arbitrary direction (neither axial nor diagonal), while
we have illustrated the interesting differences between propagating on and
off of lattice directions. A qualitative explanation for some of these
features was provided by an analytical approximation for the 2D
Peierls-Nabarro potential.

It may be interesting to extend this type of 
examination to other 1D and, especially, 2D
models, where mobile solitons may be expected, such as systems with
competing nonlinearities (the cubic-quintic model \cite{ricardo}, or the
Salerno model with competing on-site and inter-site cubic terms \cite%
{Zaragoza}) and saturable nonlinearity \cite{vicencio}, where one may expect
significant potential for genuine traveling \cite{Yugo,alan}. While herein
the mobility of solutions for realistic physical purposes in finite lattices
was examined, the existence of genuinely traveling such solutions is also
an interesting computational \cite{alan} and mathematical \cite{DEP} problem.

PGK gratefully acknowledges support from NSF-DMS-0204585, NSF-CAREER. RCG
and PGK also acknowledge support from NSF-DMS-0505663.

\end{document}